\documentclass[aps,prd,notitlepage,tightenlines,nofootinbib,twocolumn,superscriptaddress]{revtex4-2}

\usepackage{multirow}

\usepackage{amsmath}
\usepackage{amssymb,amsthm}
\usepackage{mathtools}
\usepackage{mathrsfs}
\usepackage{relsize}
\usepackage{bm}
\allowdisplaybreaks[4]  % allow equations break into several pages
\usepackage{epsfig,graphics,graphicx}
\usepackage{color,xcolor}
\usepackage{subfigure}
\usepackage{array}
\usepackage{ragged2e}
\usepackage{lineno}
\usepackage{natbib}
\usepackage{hyperref}
\hypersetup{colorlinks=true,
            linkcolor=blue,
            anchorcolor=blue,
            citecolor=magenta,
            filecolor=blue,
            urlcolor=blue,
            bookmarksnumbered=true,
            pdfview=FitB
}
\usepackage{enumitem}
\usepackage{ulem}
\usepackage{listings}
\usepackage{soul}
\usepackage{cancel}

\colorlet{darkgreen}{green!50!black}
\colorlet{brightyellow}{yellow!75!red}
\colorlet{orange}{red!50!yellow}
\colorlet{darkgray}{gray!50!black}
\colorlet{darkred}{red!50!black}

\definecolor{XHCmathematicaTitleRed}{RGB}{204,11,2}

\def\dd{{\mathrm{d}}}

\newcommand{\half}[1][1] {\mathsmaller{\frac{#1}{2}}}

\makeatletter
\newcommand*{\transpose}{%
  {\mathpalette\@transpose{}}%
}
\newcommand*{\@transpose}[2]{%
  \raisebox{\depth}{$\m@th#1\intercal$}%
}
\makeatother

\begin{document}

\title{Convergence in charmonium structure: light-front wave functions from basis light-front quantization and Dyson-Schwinger equations}

\author{Xianghui Cao}
\affiliation{Department of Modern Physics, University of Science \& Technology of China, Hefei 230026, China}

\author{Yang Li}
%\thanks{Corresponding author}
%\email{leeyoung1987@ustc.edu.cn}
\affiliation{Department of Modern Physics, University of Science \& Technology of China, Hefei 230026, China}
\affiliation{Anhui Center for Fundamental Sciences in Theoretical Physics, Hefei, 230026, China}

\author{Chao Shi}
\affiliation{Department of Nuclear Science and Technology, Nanjing University of Aeronautics and Astronautics, Nanjing 210016, China}

\author{James P. Vary}
%\email{jvary@ustc.edu.cn}
\affiliation{Department of Physics and Astronomy, Iowa State University, Ames, IA 50011, USA}

\author{Qun Wang}
\affiliation{Department of Modern Physics, University of Science \& Technology of China, Hefei 230026, China}
\affiliation{Anhui Center for Fundamental Sciences in Theoretical Physics, Hefei, 230026, China}
\affiliation{School of Mechanics and Physics, Anhui University of Science and Technology, Huainan, Anhui 232001, China}

\date{\today}

\begin{abstract}
We present a systematic comparison of charmonium light-front wave functions obtained through two complementary non-perturbative approaches: Basis Light-Front Quantization (BLFQ) and Dyson-Schwinger Equations (DSE). Key observables include the charge form factor, gravitational form factors, light-cone distribution amplitudes, decay constants, and two-photon transition form factors. Despite their distinct theoretical foundations and model parameters, the predictions from BLFQ and DSE exhibit remarkable agreement across all observables. This convergence validates both frameworks for studying charmonium structure and highlights the complementary strengths of Hamiltonian-based --- BLFQ --- and Lagrangian-based --- DSE --- methods in addressing non-perturbative QCD.
\end{abstract}

\maketitle

The charmonium system, bound states of charm and anti-charm quarks, serves as a critical laboratory for exploring non-perturbative quantum chromodynamics (QCD) \cite{Brambilla:2010cs}. It is relatively simple compared to light hadrons -- owing to the heavy quark mass, making it an ideal testbed for theoretical frameworks that bridge perturbative and non-perturbative physics.

In experimental measurements, charmonia are precise probes for extracting information from the system due to their distinct signals \cite{Brambilla:2014jmp}. For instance, the vector charmonium $J/\psi$ can be used to extract the generalized parton distribution of the nucleon in deeply virtual meson production (DVMP) and the gluon distribution in diffractive vector meson production \cite{Favart:2015umi, Kovchegov:2012mbw}. The P-wave charmonia $\chi_{c0}$ and $\chi_{c2}$ are sensitive to pomeron and odderon exchanges in $ep$ collisions and in ultra-peripheral heavy-ion collisions \cite{Ivanov:2004ax}. Additionally, the evolution of charmonia in the quark-gluon plasma (QGP) serves as a thermometer for hot quark matter \cite{Rothkopf:2019ipj}.
All of these applications hinge on a comprehensive understanding of charmonium structure, particularly in the high-energy limit, i.e., on the light front ($ct + z = 0$). Such information is fully encoded within the light-front wave functions (LFWFs), which describe hadron structures in terms of partonic degrees of freedom, enabling direct computation of observables such as form factors, distribution amplitudes, and parton distribution functions (PDFs) \cite{Lepage:1980fj}.

In recent years, there has been intense interest to access the LFWFs directly from QCD \cite{Gross:2022hyw}.
Two non-perturbative approaches -- basis light-front quantization (BLFQ) and Dyson-Schwinger equations (DSE) -- have emerged as powerful tools for investigating charmonium \cite{Gross:2022hyw}.
BLFQ diagonalizes the light-front QCD Hamiltonian in a symmetry-preserving basis, providing direct access to hadron spectra and wave functions \cite{Vary:2009gt, Wiecki:2014ola}.
BLFQ with effective interactions inspired by holographic QCD has been used to investigate a variety of systems, such as heavy quarkonia \cite{Li:2015zda}, heavy-light mesons \cite{Tang:2018myz, Tang:2019gvn}, light mesons \cite{Qian:2020utg, Jia:2018ary, Lan:2021wok}, nucleons \cite{Mondal:2019jdg, Xu:2021wwj, Xu:2024sjt} and tetraquarks \cite{Kuang:2022vdy}, and to a number of observables including mass spectra \cite{Li:2017mlw}, form factors \cite{Mondal:2019jdg, Nair:2024fit, Xu:2024hfx, Hu:2024edc}, radiative transitions \cite{Li:2018uif, Tang:2020org, Li:2021ejv, Wang:2023nhb}, parton distributions \cite{Lan:2019vui, Xu:2021wwj, Adhikari:2018umb, Adhikari:2021jrh, Liu:2022fvl, Zhang:2023xfe, Lin:2023ezw, Kaur:2023lun, Lin:2024ijo, Liu:2024umn, Lan:2024ais}, spin structures \cite{Xu:2022abw, Xu:2023nqv}, and transverse momentum distributions \cite{Hu:2022ctr, Zhu:2023lst, Zhu:2024awq}.

In contrast, DSE solves QCD’s Green’s functions in the continuum, emphasizing symmetry and Lorentz covariance \cite{Dyson:1949ha, Schwinger:1951ex, Schwinger:1951hq}. The Maris-Tandy (MT) model combined with Rainbow-Ladder (RL) truncation exposes the key role of the dynamical chiral symmetry breaking (D$\chi$SB) in describing the structure of the pion as a Goldstone boson \cite{Maris:2003vk}. This approach, including the subsequent improvements of the MT model \cite{Qin:2011dd}, has been successfully applied to investigate the meson spectrum, baryon spectrum, decay constants, electromagnetic form factors, axial form factors, gravitational form factors and radiative widths \cite{Maris:2003vk, Roberts:2007ji, Cloet:2013jya, Eichmann:2016yit}. See Ref.~\cite{Eichmann:2016yit} for a recent review.

Recently, it was shown that $k_\perp$-dependent moments can be used to project the covariant Bethe-Salpeter amplitudes (BSA) to the light cone \cite{Shi:2018zqd, Shi:2020pqe}. LFWFs extracted from this method have been used to investigate a number of partonic observables, including parton distribution amplitudes, 1-dimensional and 3-dimensional parton distribution functions \cite{Shi:2021taf, Shi:2021nvg, Shi:2022erw, Shi:2023oll, Kou:2023ady, Shi:2024laj}.

\begin{figure}
  \centering
  \includegraphics[width=0.6\columnwidth]{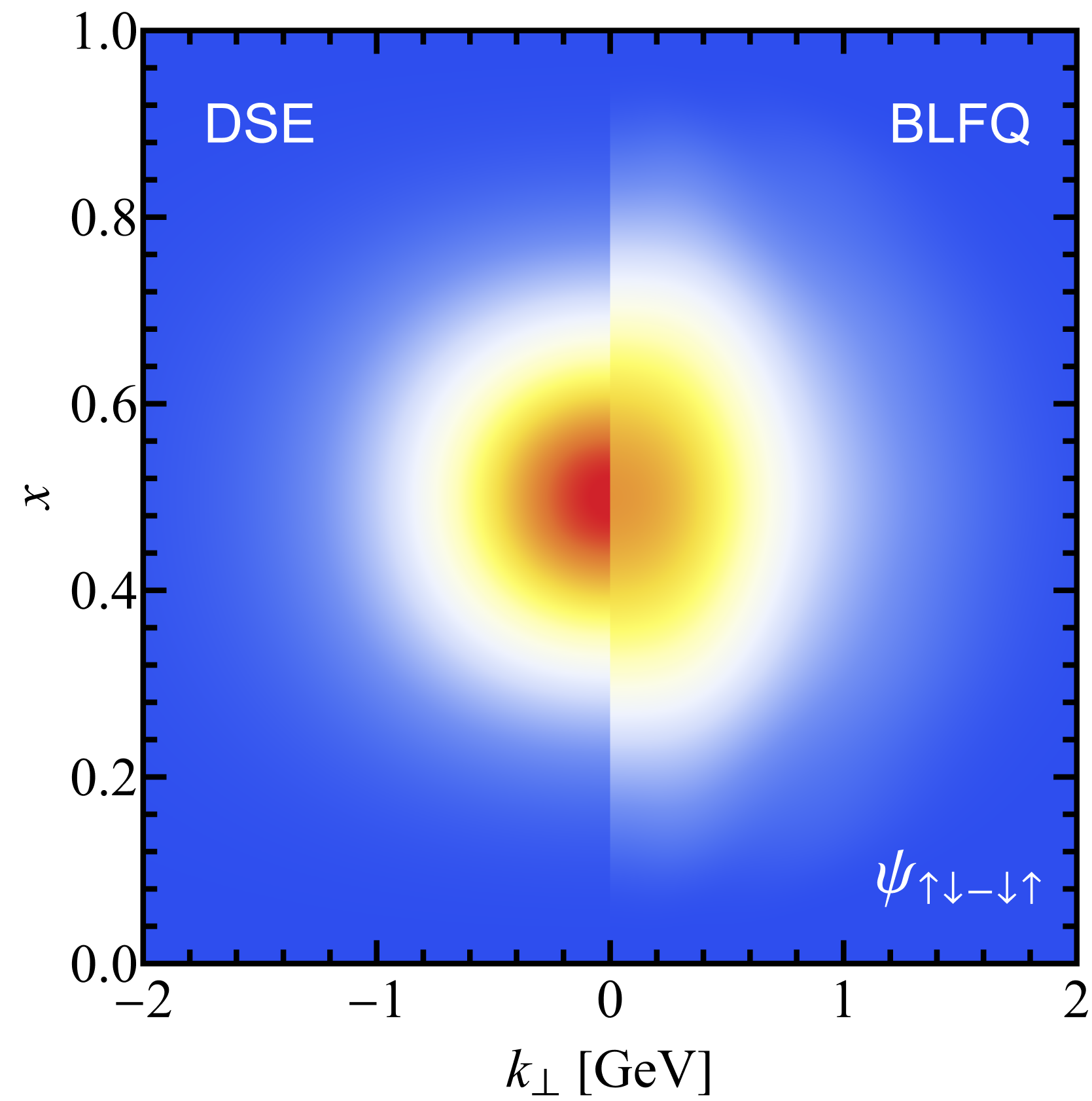}
  \includegraphics[width=0.6\columnwidth]{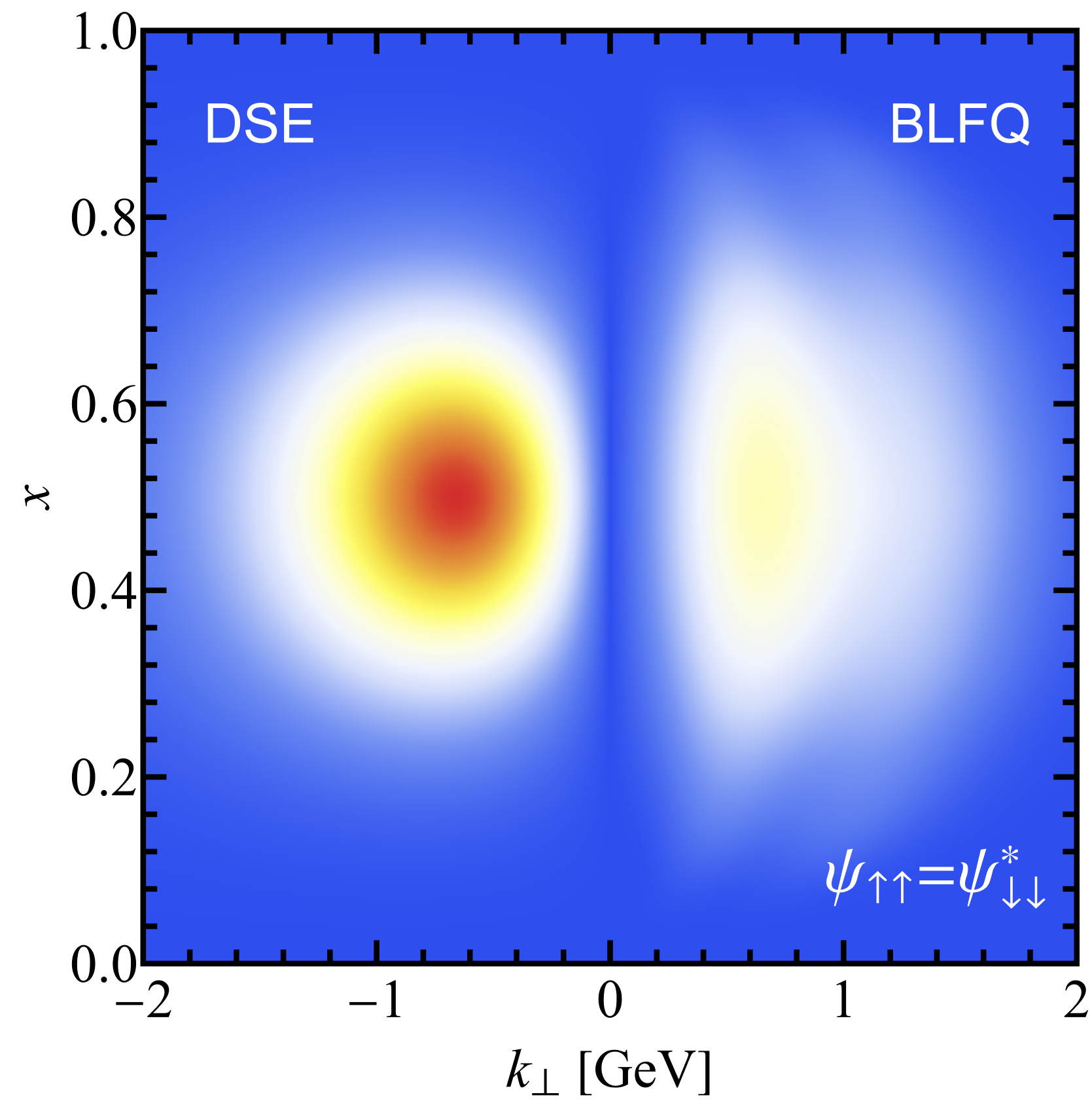}
  \caption{Comparison of the light-front wave functions for the ground-state charmonium $\eta_c$ computed from basis light-front quantization (BLFQ) and from Dyson-Schwinger equations (DSE). The negative $k_\perp$ region shows the DSE results. The positive $k_\perp$ region shows the BLFQ results with $N_\text{max} = 8$. The spin singlet component of the LFWF is defined as $\psi_{\uparrow\downarrow-\downarrow\uparrow} = (\psi_{\uparrow\downarrow} - \psi_{\downarrow\uparrow})/\sqrt{2}$.}
  \label{fig:etac_LFWF}
\end{figure}
Since both methods have independently reproduced key charmonium properties, despite their conceptual differences,  there is a strong motivation to scrutinize the underlying LFWFs -- the fundamental entities encoding hadron structure -- obtained from these two approaches through common observables. Similar comparisons were performed in the literature between various approaches and for various systems, e.g. Refs.~\cite{Leitao:2017esb, Lan:2024ais}.
In this work, we address this question by comparing BLFQ and DSE predictions across five sets of structural probes: (\textit{i}) charge form factor, (\textit{ii}) gravitational form factors, (\textit{iii}) light-cone distribution amplitudes, (\textit{iv}) decay constants, and (\textit{v}) two-photon transition form factors.
The same LFWFs -- obtained from BLFQ and DSE respectively -- are used for each set of observable, to avoid additional phenomenological modifications.
The observed agreement, despite differing model parameters (e.g., regulator scales, interaction kernels), highlights the universality of QCD-driven charmonium properties and strengthens confidence in both approaches for future studies of more challenging systems and more complicated observables.

The remainder of this article is structured as follows. Section~\ref{sec:lfwfs} briefly reviews the theoretical framework and properties of the light-front wave functions. Sec.~\ref{sec:FF} presents the hadronic form factors. Sec.~\ref{sec:LCDA} compares the light-cone distribution amplitudes, and the two-photon transition form factors closely related to the LCDA.
Finally, we summarize in Sec.~\ref{sec:summary}.

\section{Light-front wave functions}\label{sec:lfwfs}

The LFWFs are the amplitudes of the hadronic state vector in Fock space at fixed light-front time $x^+$ \cite{Brodsky:1997de}. Specifically, the charmonium hadronic state vector on the light front can be written as,
\begin{multline}\label{eqn:hadronic_state_vector}
|\psi_h(p, j, \sigma)\rangle = \sum_{s, \bar s} \int_0^1\frac{\dd x}{2x(1-x)}\int\frac{\dd^2k_\perp}{(2\pi)^3} 
\psi_{s\bar s/h}^{\sigma}(x, \vec k_\perp) \\
\times \frac{1}{\sqrt{N_c}}\sum_i b^\dagger_{si}(p_1) d^\dagger_{\bar s i}(p_2) |0\rangle + \cdots
\end{multline}
where, $p$ is the 4-momentum of the particle; $j$ and $\sigma$ are the particle's total angular momentum and magnetic projection, respectively. The coefficients $\psi_{s\bar{s}/h}^\sigma(x,\vec{k}_\perp)$ are the valence sector LFWFs, with $s(\bar{s})$ representing the spin of the quark (antiquark). $x = p^+_1/p^+$ is the longitudinal momentum fraction, $\vec k_\perp = \vec p_{1\perp} - x\vec p_\perp$ is the relative transverse momentum and $p_1^2 = p_2^2 = m_q^2$. The ellipsis represents the non-valence Fock contributions. The quark and antiquark creation operators satisfy the anticommutation relations:
\begin{multline}
\{b_{si}(p), b_{s^\prime i^\prime}^\dagger(p^\prime)\} = \{d_{si}(p), d_{s^\prime i^\prime}^\dagger(p^\prime)\}\\
=2 p^+(2\pi)^3 \delta^{(3)}(p - p^\prime) \delta_{ss^\prime}\delta_{ii^\prime}.
\end{multline}
Here, $i$ denotes the color charge of quark and antiquark, and the total color number is $N_c=3$.

In principle, the hadronic state vector, hence the LFWFs, can be directly obtained from diagonalizing the light-cone Hamiltonian operator $H_\textsc{lc} = P^\mu P_\mu \equiv P^+ P^- - \vec P_\perp^2$ \cite{Brodsky:1997de}, 
 \begin{equation}\label{eqn:LC_eigenvalue_equation}
H_\textsc{lc}  |\psi_h(p)\rangle =  M^2_h |\psi_h(p)\rangle.
\end{equation}
This is the starting point of BLFQ. On the other hand, from Eq.~(\ref{eqn:hadronic_state_vector}) the LFWFs can also be written as the hadronic amplitude of the appropriate bi-local operators \cite{Ji:2003yj}. For example, the spin-flip LFWF of a pseudoscalar meson $P$ can be expressed as,
\begin{multline}
\psi_{\uparrow\downarrow-\downarrow\uparrow/P}(x, \vec k_\perp) = \frac{1}{2p^+}\int \dd^3 z \, e^{\half[i] xp\cdot z - i\vec k_\perp \cdot \vec z_\perp} \\
\times \langle 0 | \bar \psi(0)\gamma^+\gamma_5 \psi(z)|P(p)\rangle_{z^+=0},\label{eq:LFWF-Projection}
\end{multline}
where $z$ is the four-vector in Minkowski space. In this way, the LFWFs can also be obtained from the covariant Bethe-Salpeter amplitudes.

In both the BLFQ and the DSE approaches, the symmetries of individual LFWFs are fully retained. For example, the pseudoscalar LFWF consists of two independent structures \cite{Leitner:2010nx},
\begin{equation}
\psi_{s\bar s/P}(x, \vec{k}_\perp) = \bar v_{\bar s}(p_2)\Big[
\gamma^5 \phi_1(x, \vec{k}_\perp^2) + \frac{\gamma^+ \gamma^5 }{p^+}\phi_2(x, \vec{k}_\perp^2)\Big] u_{s}(p_1).
\end{equation}
The second term appears because the LFWFs in general depend on the orientation of light front $\omega$, which is a null vector ($\omega^2 = 0$, and $\omega\cdot \gamma = \gamma^+$). In some phenomenological applications, the second term is dropped \cite{Jaus:1999zv, Cheng:2003sm, Wang:2008xt, Verma:2011yw, Choi:2004ww, Wang:2024cyi, Arifi:2024mff}. However, we have shown that this term plays an important role in chiral symmetry breaking \cite{Li:2022mlg}. In both BLFQ and DSE, the maximal kinematical symmetries of the LFWFs are retained and both approaches produce two independent structures.
Figure~\ref{fig:etac_LFWF} compares the valence LFWFs of $\eta_c$ as obtained from BLFQ and from DSE.
The LFWFs of BLFQ and DSE are accessible from data repository \cite{Li:charmoniumData} and \cite{Shi:charmoniumData}, respectively.
For both spin-singlet $\psi_{\uparrow\downarrow-\downarrow\uparrow}$ and spin-triplet $\psi_{\uparrow\uparrow}/\psi_{\downarrow\downarrow}$ components, the LFWFs from BLFQ are broader than DSE in the transverse momentum ($k_\perp$) direction and longitudinal ($x$) direction.
However, the normalizations of each of the spin configuration are qualitatively similar in these two frameworks.
The normalization for each spin configuration is defined as
\begin{gather}
  I_{s\bar{s}} = \int_0^1 \frac{\dd x}{2x(1-x)} \int \frac{\dd^2 k_\perp}{(2\pi)^3} |\psi_{s\bar{s}/h}^\sigma(x, \vec{k}_\perp)|^2.
\end{gather}
For BLFQ, we find $I_{\uparrow\uparrow} + I_{\downarrow\downarrow} = 0.074$ and $I_{\uparrow\downarrow-\downarrow\uparrow} = 0.926$, whereas the DSE LFWFs give $I_{\uparrow\uparrow} + I_{\downarrow\downarrow} = 0.140$ and $I_{\uparrow\downarrow - \downarrow\uparrow} = 0.860$. In both frameworks, the spin-singlet component is dominant.

A comment is in order here. The valence Fock sector LFWFs projected from the DSE are not automatically normalized to 1 due to contributions from high Fock states in the BSA. However, for heavy quarkonia studied in this paper, the valence Fock sector contributes more than 90\% of the total normalization \cite{Shi:2021nvg}. To be consistent with the BLFQ approach, we have normalized the DSE LFWFs to 1. All subsequent calculations will use these normalized LFWFs. 

In practical calculations, the LFWFs within the BLFQ are expanded in a basis of 2D harmonic oscillator functions in the transverse directions and Jacobi functions in the longitudinal momentum fraction $x$ \cite{Li:2017mlw}. To render the basis finite for numerical implementation, the quantum numbers are truncated according to $2n + |m| + 1 \leq N_\text{max}$ and $0\leq l \leq L_\text{max}$,  where $n, m$ denote the radial and angular quantum numbers of the 2D harmonic oscillator modes and $l$ labels the order of the Jacobi functions. This $N_\text{max}$-truncation naturally introduces both UV and infrared IR cutoffs: $\Lambda_\text{UV} = b\sqrt{N_\text{max}}$, $\lambda_\text{IR} = b / \sqrt{N_\text{max}}$, where $b$ is the oscillator basis scale. Meanwhile, $L_\text{max}$ controls the resolution in the longitudinal momentum fraction $\Delta x \approx L_\text{max}^{-1}$. In contrast, the DSE approach adopts an explicit UV cutoff $\Lambda$ on the Euclidean momentum $k_E$ of the Green’s functions. In practice, $\Lambda$ is chosen sufficiently large to ensure convergence while minimizing regulator artifacts \cite{Shi:2021nvg}.

\section{Form factors and hadronic densities}\label{sec:FF}

Hadronic form factors are defined from the hadronic matrix elements of local current operators. In LFWF representation, they are related to the Fourier transform of the one-body hadronic densities. In this section, we focus on the charge form factor and the gravitational form factor.

The charge form factor $F(Q^2)$ is defined from the hadronic matrix elements of the electromagnetic vector current $J^\mu(x)$,
\begin{equation}
\langle \psi(p')|J^\mu(0)|\psi(p)\rangle = 2P^\mu F(Q^2),
\end{equation}
where, $P = (p'+p)/2$, $q = p'-p$ and $Q^2 = -q^2$.
In light-front dynamics, $F(Q^2)$ can be interpreted as the Fourier transform of the transverse charge density \cite{Miller:2010nz}:
\begin{equation}
\rho(r_\perp) = \int \frac{\dd^2 q_\perp}{(2\pi)^2} e^{-i\vec q_\perp \cdot \vec r_\perp} F(q_\perp^2).
\end{equation}
Unlike the Breit frame density, the above charge distribution is frame independent and is a genuine distribution \cite{Miller:2018ybm, Jaffe:2020ebz}.
The Drell-Yan-West formula expresses the transverse charge density as a one-body density (OBD) using LFWFs \cite{Drell:1969km, West:1970av, Brodsky:1998hn},
\begin{equation}
\rho(r_\perp) = \Big\langle \sum_i e_i \delta^2(r_\perp - r_{i\perp}) \Big\rangle
\end{equation}
where, the $\langle \cdots \rangle$ represents the quantum average:
\begin{multline}
\langle O \rangle = \sum_{n} \frac{1}{S_n}\sum_{s_i}\prod_{i}^n  \int \frac{\dd x_i}{4\pi x_i}\int \dd^2 r_{i\perp}
4\pi\delta(\sum_j x_j-1) \\
\times \delta^2(\sum_j x_j \vec r_{j\perp}) \big|\widetilde\psi_n(\{x_i, \vec r_{i\perp}, s_i\})\big|^2 O(\{x_i, \vec r_{i\perp}, s_i\}).
\end{multline}
Here, $S_n$ is the symmetry factor. $\psi_n(\{x_i, \vec r_{i\perp}, s_i\})$ is the transverse coordinate space wave function defined as, 
\begin{multline}\label{eqn:intrinsic_coordinate_lfwf}
\widetilde\psi_n(\{x_i, \vec r_{i\perp}, s_i\}) = \prod_{i=1}^{n} \int \frac{\dd^2k_{i\perp}}{(2\pi)^2} (2\pi)^2\delta^2(\sum_i k_{i\perp}) \\
\times e^{-i\sum_i\vec k_{i\perp}\cdot \vec r_{i\perp}}\psi_n(\{x_i, \vec k_{i\perp}, s_i\}). 
\end{multline}

\begin{figure}
\centering
\includegraphics[width=0.85\columnwidth]{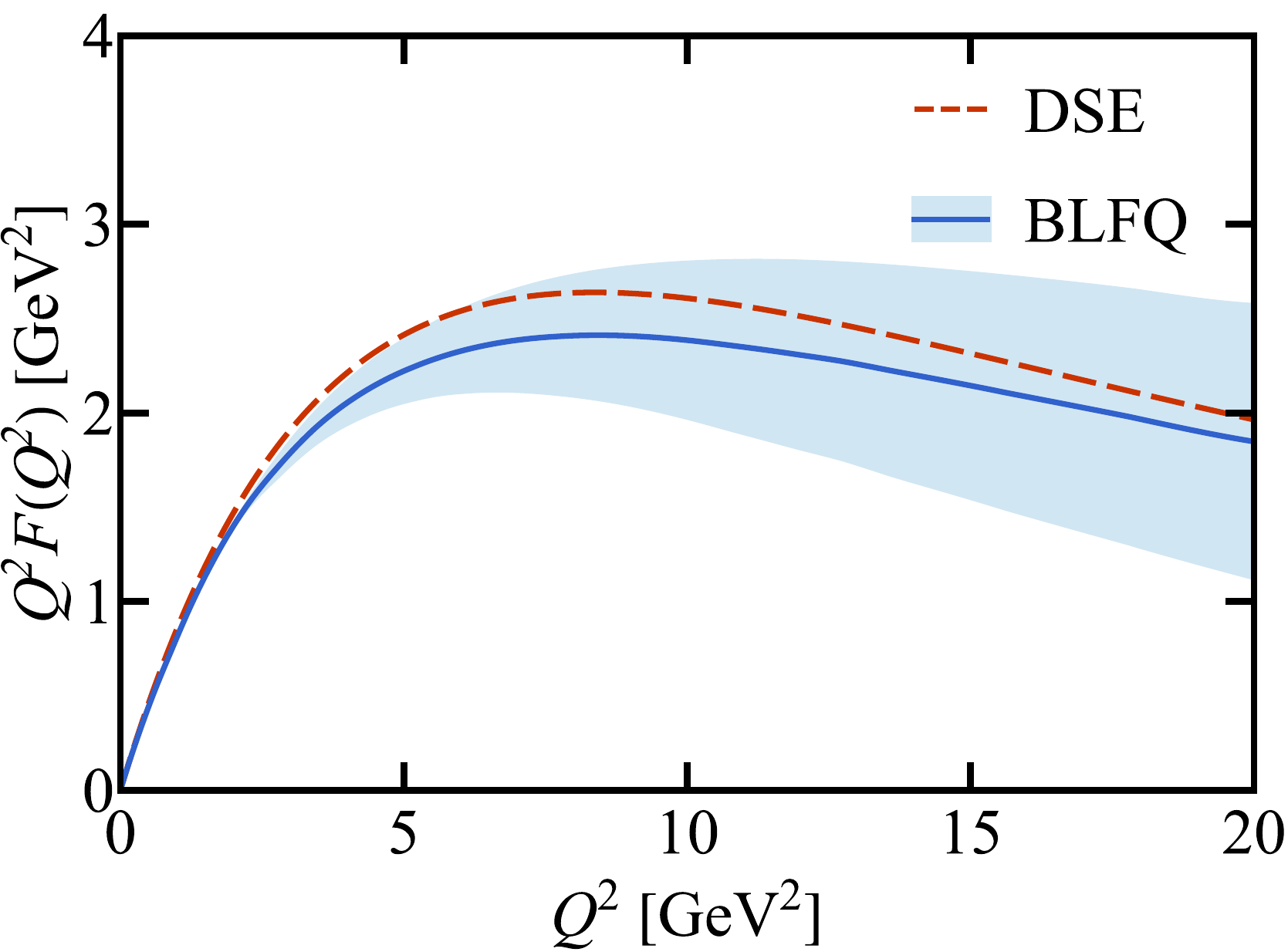}
\caption{(Colors online) Comparison of the (fictitious) charge form factor $F(Q^2)$ for the ground-state charmonium $\eta_c$ computed using LFWFs from BLFQ and from DSE. The dashed line is the DSE result. The solid line is the BLFQ result with $N_\mathrm{max} = 8$. And the band represents the uncertainty associated with the basis sensitivity in BLFQ as computed from the difference between $N_\mathrm{max} = 8$ and $N_\mathrm{max} = 16$ results. See texts for more details. }
\label{fig:charge_FF}
\end{figure}

For charmonium, the physical charge form factor vanishes due to the charge conjugation symmetry. Nevertheless, it is custom to define a fictitious charge form factor, where the photon couples to the quark and antiquark differently, similar to the charged pion $\pi^{\pm}$ and $B_c$, while the quark masses remain unchanged $m_q = m_{\bar q} = m_c$ \cite{Dudek:2006ej}. Figure~\ref{fig:charge_FF} compares the charge form factor computed using LFWFs from BLFQ and from DSE.  Following our previous analyses for dilepton, diphoton and radiative transitions, hereafter we adopt $N_\mathrm{max} = 8$ LFWFs for central values of BLFQ results, which corresponds to a UV scale $\Lambda_\text{UV} = 2.8\,\text{GeV} \approx M_{c\bar c}$, and quote the difference between $N_\mathrm{max} = 8$ and $N_\mathrm{max} = 16$ as the basis sensitivity \cite{Li:2017mlw, Li:2021ejv, Wang:2023nhb}, where $N_\mathrm{max}$ denotes the basis truncation parameter. This choice is motivated by the competition between the needs for both a better basis resolution and a lower UV scale to suppress the radiative corrections.  The primary source of uncertainty in the DSE results stems from the RL approximation, which is difficult to estimate. Consequently, only the central values are presented here and in the subsequent calculations. As shown in Fig.~\ref{fig:charge_FF}, the BLFQ and DSE results are in good agreement with each other including at high $Q^2$ ($Q^2\gtrsim 10\,\mathrm{GeV^2}$).

The gravitational form factors are defined via the hadronic matrix elements of the energy-momentum tensor $T^{\mu\nu}(x)$,
\begin{multline}
\langle \psi(p')|T^{\mu\nu}(0)|\psi(p)\rangle = 2P^\mu P^\nu A(Q^2) \\
+ \frac{1}{2}(q^\mu q^\nu - g^{\mu\nu} q^2)D(Q^2).
\end{multline}
In light-front dynamics, $A$ and $D$ are related to the Fourier transform of the transverse flow density $\mathcal A(r_\perp)$, and transverse shear density $ \mathcal T^{12}(r_\perp) = \mathcal S(r_\perp)/(2P^+)$
\begin{align}
\mathcal A(r_\perp) =\,& \int \frac{\dd^2 q_\perp}{(2\pi)^2} e^{-i\vec q_\perp \cdot \vec r_\perp} A(q_\perp^2), \\
\mathcal S(r_\perp) =\,&  \int \frac{\dd^2 q_\perp}{(2\pi)^2} e^{-i\vec q_\perp \cdot \vec r_\perp} \frac{1}{2} q^1_\perp q^2_\perp D(q^2_\perp).
\end{align}
The Brodsky-Hwang-Ma-Schmidt formula gives the transverse flow density as an OBD using LFWFs \cite{Brodsky:2000ii}, 
\begin{equation}
\mathcal A(r_\perp) = \Big\langle \sum_i x_i \delta^2(r_\perp - r_{i\perp}) \Big\rangle.
\end{equation}
Recently, some of us provided the LFWF representation of the transverse shear density as an OBD \cite{Cao:2023ohj, Cao:2024rul, Cao:2024fto},
\begin{equation}\label{eqn:t12}
\mathcal S(r_\perp) = \Big\langle \sum_i \frac{i\tensor\nabla^1_{i\perp}i\tensor\nabla^2_{i\perp} - i\nabla^1_\perp i\nabla^2_\perp}{2x_i} \delta^2(r_\perp - r_{i\perp})\Big\rangle
\end{equation}
where, $\tensor\nabla_\perp$ only acts on the suppressed wave functions and $f \tensor\nabla g = f \nabla g - (\nabla f) g$.

\begin{figure}
\centering
\includegraphics[width=0.85\columnwidth]{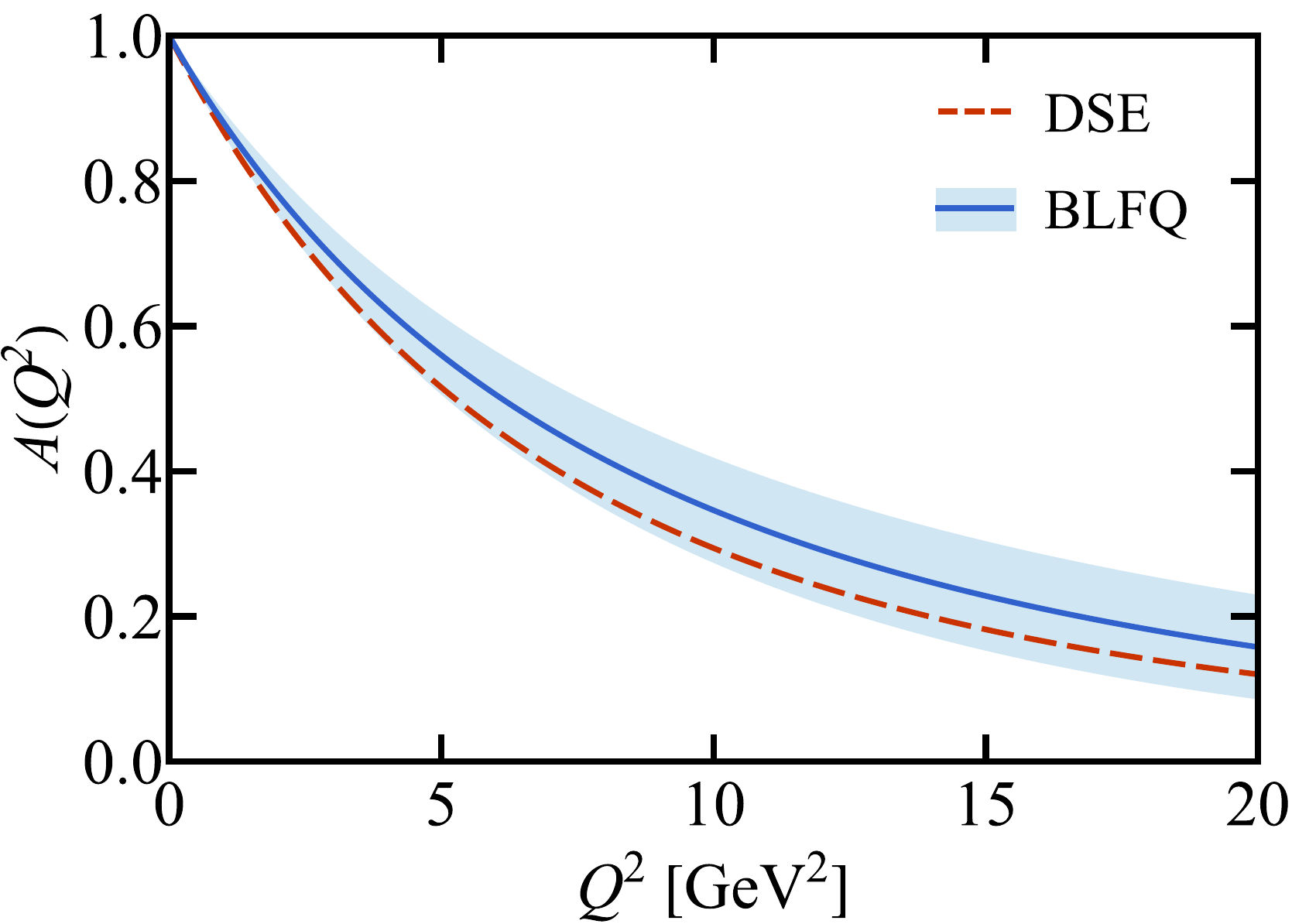}
\includegraphics[width=0.85\columnwidth]{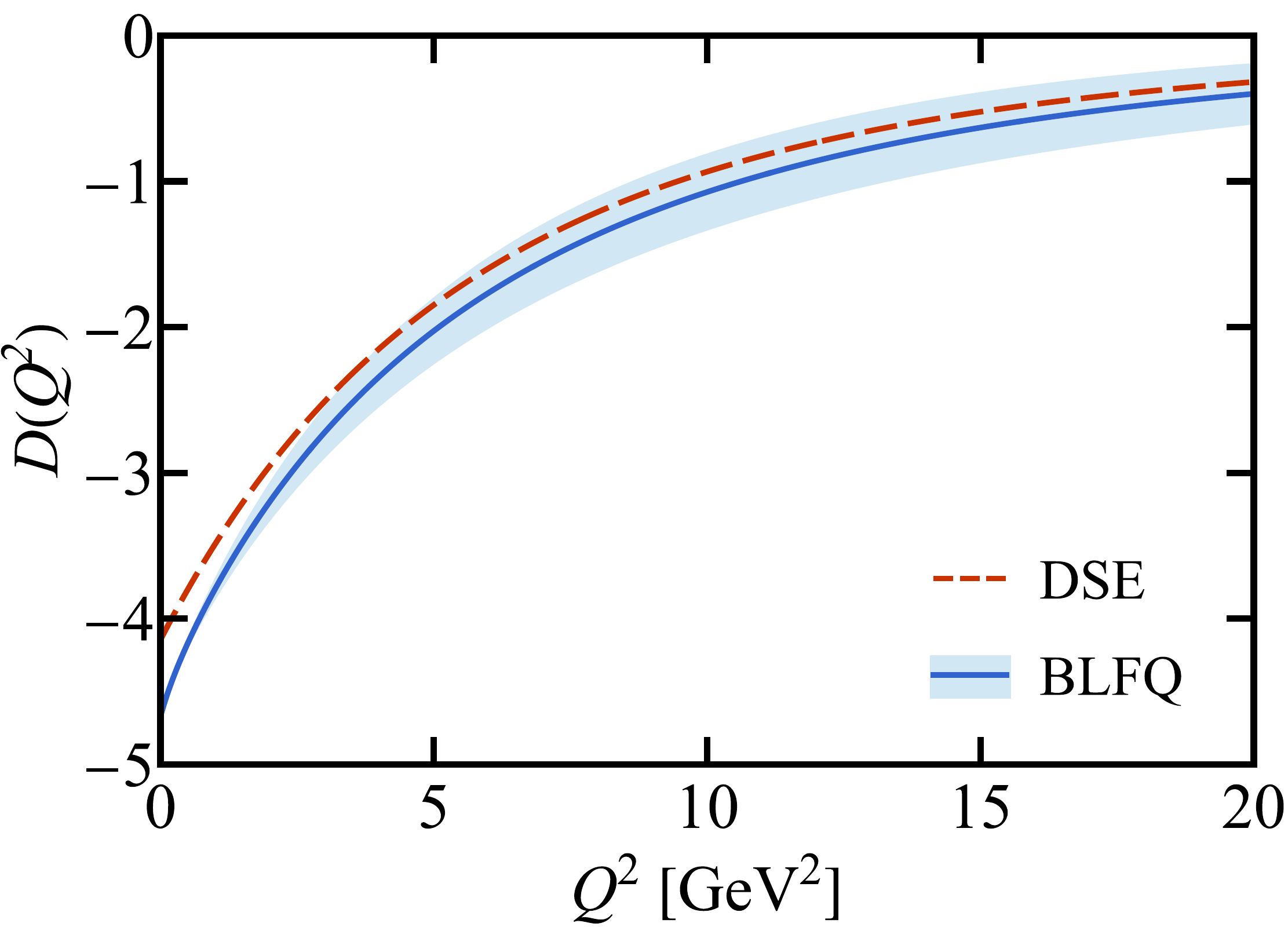}
\caption{(Colors online) Comparison of the gravitational form factors $A(Q^2)$ and $D(Q^2)$ for the ground-state charmonium $\eta_c$ computed using LFWFs from BLFQ ($N_\text{max} = 8$, solid lines) and from DSE (dashed lines). The uncertainty bands are computed from the difference between $N_\mathrm{max} = 8$ and $N_\mathrm{max} = 16$ results in BLFQ. See texts for more details.}
\label{fig:GFF}
\end{figure}

Figure~\ref{fig:GFF} compares the gravitational form factors $A(Q^2)$ and $D(Q^2)$ computed using LFWFs from BLFQ and from DSE. As one can see, the BLFQ and DSE results are in good agreement. The $D$-term, defined as $D \equiv D(0)$, is known to be sensitive to the interaction.  In Ref.~\cite{Sultan:2024hep}, the authors argue that the $D$-term of the pseudoscalar meson should lie within $(-1, -1/3)$, which are the values in the chiral limit and the infinite mass limit respectively. Using DSE with a contact interaction, they obtain $D = -0.58$ for $\eta_c$. In contrast, the values extracted from BLFQ and from DSE LFWFs in this work are $D_\textsc{blfq} = -4.6$ and $D_\textsc{dse} = -4.1$.

\section{Light-cone distribution amplitudes and radiative transitions}\label{sec:LCDA}

\begin{figure}
\centering
\includegraphics[width=0.85\columnwidth]{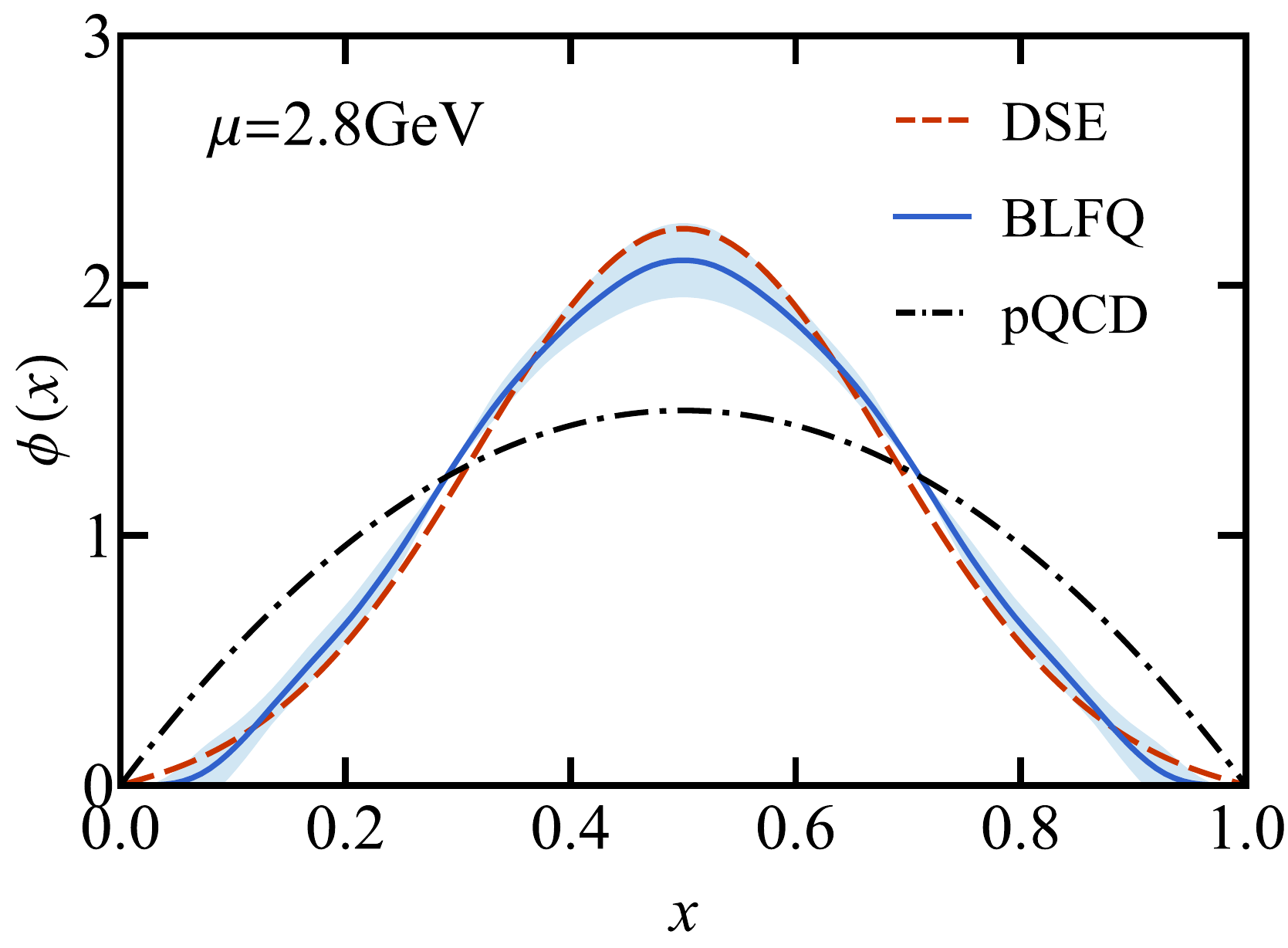}
\caption{(Colors online) Comparison of the leading-twist LCDA of $\eta_c$ as obtained from BLFQ ($N_\text{max} = 8$, solid lines) and from DSE (dashed lines). The uncertainty bands are computed from the difference between $N_\text{max} = 8$ and $N_\text{max} = 16$ results in BLFQ. The dot-dashed lines correspond to the perturbative QCD (pQCD) asymptotic form $6x(1-x)$ \cite{Lepage:1980fj}. The DSE results are evolved from $\mu_0 = 2.6\,\mathrm{GeV}$ to $\mu = 2.8\,\mathrm{GeV}$ to match the BLFQ energy scale.}
\label{fig:DA}
\end{figure}

While the form factors probe the transverse structures of the system, the light-cone distribution amplitudes (LCDA) describe its longitudinal structure. LCDAs are defined as the bi-local light-like vacuum-to-hadron transition amplitudes. For pseudoscalar $P$, the leading-twist LCDA $\phi_P(x, \mu)$ is,
\begin{multline}
\langle 0 | \bar \psi(-\half z) \gamma^+\gamma_5 \psi(+\half z)|P(p)\rangle_\mu \\
= if_P p^+\int_0^1 \dd x e^{\half[i] xp^+z^-} \phi_P(x, \mu).
\end{multline}
where $f_P$ is the decay constant, and $z$ is the same as in Eq.~\eqref{eq:LFWF-Projection}. Using LFWFs, the above LCDA can be expressed as,
\begin{multline}
\frac{f_P}{2\sqrt{2N_c}} \phi_P(x, \mu) = \\
\frac{1}{2\sqrt{x(1-x)}} \int^{\mu^2}\frac{\dd^2k_\perp}{(2\pi)^3}\psi_{\uparrow\downarrow-\downarrow\uparrow/P}(x, \vec k_\perp).
\label{eq:LCDA_P}
\end{multline}
Figure~\ref{fig:DA} compares the leading-twist LCDA of $\eta_c$ as obtained from BLFQ and from DSE and evolved to $\mu = 2.8\,\mathrm{GeV}$. The DSE result appears narrower than the BLFQ result. However, both are broad enough to distinguish from the nonrelativistic limit. The decay constant can be derived after integrating over $x$ in Eq. \eqref{eq:LCDA_P},
\begin{gather}
  \frac{f_P}{2\sqrt{2N_c}} = \int_0^1 \frac{\dd x}{2\sqrt{x(1-x)}} \int \frac{\dd^2 k_\perp}{(2\pi)^3} \psi_{\uparrow\downarrow-\downarrow\uparrow}(x, \vec{k}_\perp).
\end{gather}
The BLFQ LFWFs of $\eta_c$ yield a decay constant $f_P = 0.414(78)\,\mathrm{GeV}$ while the DSE LFWFs give $f_P = 0.396\,\mathrm{GeV}$. Both results are close to the Lattice QCD result $f_P = 0.395(2)\,\mathrm{GeV}$ \cite{Davies:2010ip}.

\begin{figure}
\centering
\includegraphics[width=0.85\columnwidth]{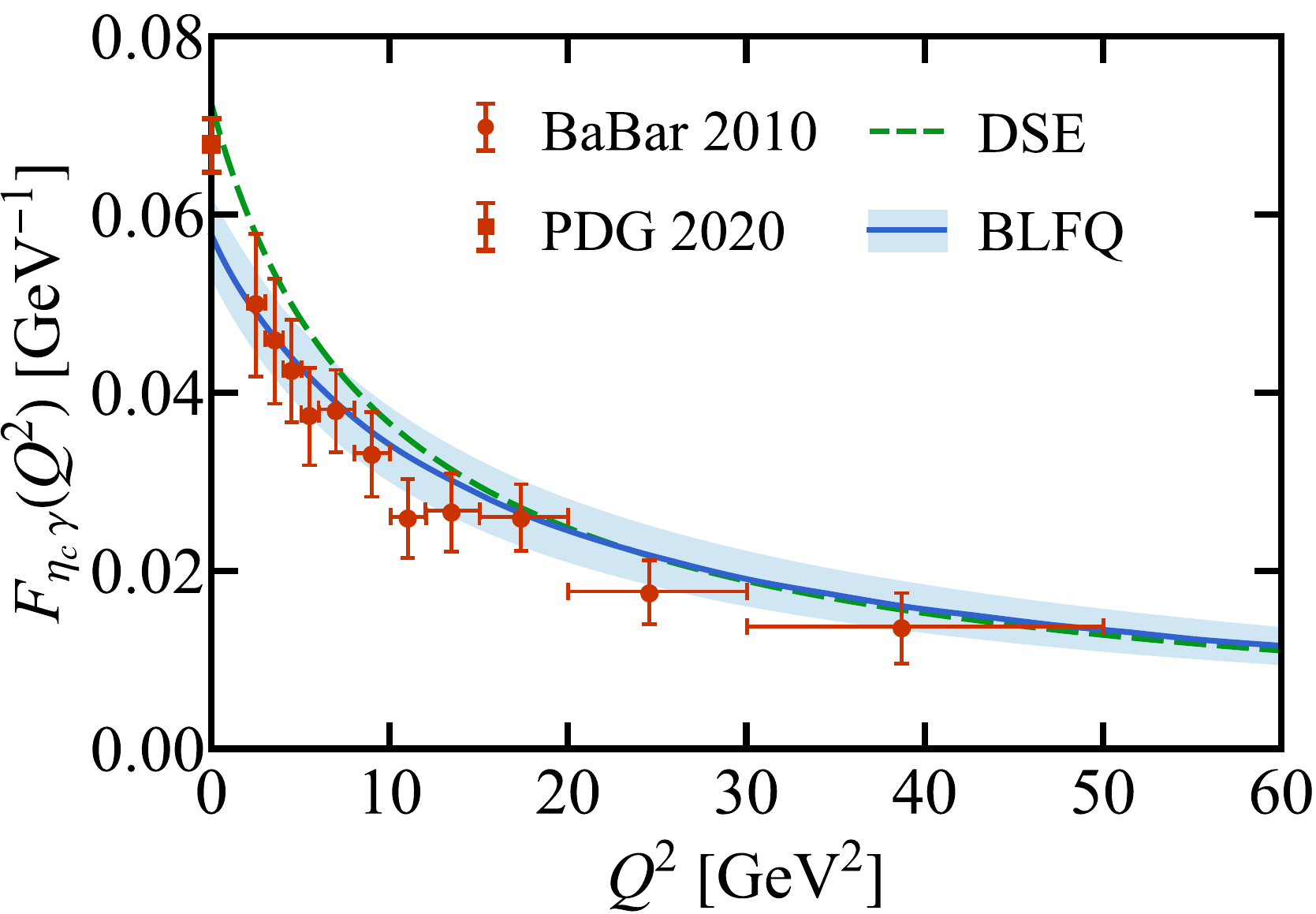}
\caption{(Colors online) Comparison of the transition form factor of $\eta_c$ as obtained from BLFQ ($N_\text{max} = 8$, solid lines) and from DSE (dashed lines). The uncertainty bands are computed from the difference between $N_\text{max} = 8$ and $N_\text{max} = 16$ results in BLFQ.}
\label{fig:TFF}
\end{figure}

The LCDA is an essential tool for describing exclusive processes in high-energy scattering $Q^2 \gg \Lambda_\textsc{qcd}^2$. One of the observables that can be described by LCDA is the two-photon transition form factor (TFF), as shown by Lepage \& Brodsky \cite{Lepage:1980fj}. On the other hand, the evaluation of the TFF can be extended to low $Q^2$ via LFWFs \cite{Lepage:1980fj, Babiarz:2019sfa, Li:2021ejv}.
Figure~\ref{fig:TFF} compares the evaluation of the TFF $F_{\eta_c\gamma}(Q^2)$ for the process $\gamma^*\gamma \to \eta_c$ using LFWFs obtained from BLFQ and from DSE. The theoretical results, without tuning of parameters, are in good agreement with \textsc{BaBar} measurements.

\section{Summary and conclusions}\label{sec:summary}

In this work, we have presented a comprehensive comparison of charmonium LFWFs computed using BLFQ and DSE approaches, evaluating five key observables that probe different aspects of hadron structure. The remarkable agreement between the two approaches, despite their independent methodologies and model parameters, underscores the robustness of their respective non-perturbative QCD predictions for heavy quarkonia. This work not only validates the BLFQ and DSE frameworks but also provides a foundation for future studies of more complex systems and processes in hadronic physics.

This work also validates the reliability of the BSA projection method in heavy quarkonia. The observed agreement demonstrates that the leading Fock sector LFWFs projected from BSA are sufficiently accurate to describe the hadron structure. Actually, direct calculation of the normalization indicates that the valence Fock sector is dominant for heavy quarkonia. However, for light mesons like the pion, higher Fock states are more important.

This study demonstrates shared features and predictive power within Hamiltonian-based and Lagrangian-based approaches toward LFWFs, providing a unified perspective on charmonium structure. The agreement also corroborates the reliability of these frameworks for future investigations of more complex systems, such as exotic hadrons, heavy-light mesons, and even nuclei. Furthermore, the consistency between BLFQ and DSE predictions strengthens the reliability of both non-perturbative QCD methods for interpreting experimental data from facilities like Jefferson Lab, the LHC, and the future electron-ion colliders.

\acknowledgments

We wish to thank X. Zhao for fruitful discussions. 
This work is supported in part by the National Natural Science Foundation of China (NSFC) under Grant No.~12375081, and by the Chinese Academy of Sciences under Grant No.~YSBR-101. X.-h.~Cao is supported by the NSFC under Grant No.~125B2111.

\section*{Data availability}

The data that support the findings of this article are openly available \cite{Li:charmoniumData, Shi:charmoniumData}.


\begin{thebibliography}{99}

%\cite{Brambilla:2010cs}
\bibitem{Brambilla:2010cs}
N.~Brambilla, S.~Eidelman, B.~K.~Heltsley, R.~Vogt, G.~T.~Bodwin, E.~Eichten, A.~D.~Frawley, A.~B.~Meyer, R.~E.~Mitchell and V.~Papadimitriou, \textit{et al.}
%``Heavy Quarkonium: Progress, Puzzles, and Opportunities,''
Eur. Phys. J. C \textbf{71}, 1534 (2011)
doi:10.1140/epjc/s10052-010-1534-9
[arXiv:1010.5827 [hep-ph]].
%1945 citations counted in INSPIRE as of 22 May 2025

%\cite{Brambilla:2014jmp}
\bibitem{Brambilla:2014jmp}
N.~Brambilla, S.~Eidelman, P.~Foka, S.~Gardner, A.~S.~Kronfeld, M.~G.~Alford, R.~Alkofer, M.~Butenschoen, T.~D.~Cohen and J.~Erdmenger, \textit{et al.}
%``QCD and Strongly Coupled Gauge Theories: Challenges and Perspectives,''
Eur. Phys. J. C \textbf{74}, no.10, 2981 (2014)
doi:10.1140/epjc/s10052-014-2981-5
[arXiv:1404.3723 [hep-ph]].
%602 citations counted in INSPIRE as of 23 May 2025

%\cite{Favart:2015umi}
\bibitem{Favart:2015umi}
L.~Favart, M.~Guidal, T.~Horn and P.~Kroll,
%``Deeply Virtual Meson Production on the nucleon,''
Eur. Phys. J. A \textbf{52}, no.6, 158 (2016)
doi:10.1140/epja/i2016-16158-2
[arXiv:1511.04535 [hep-ph]].
%90 citations counted in INSPIRE as of 15 May 2025

%\cite{Kovchegov:2012mbw}
\bibitem{Kovchegov:2012mbw}
Y.~V.~Kovchegov and E.~Levin,
%``Quantum Chromodynamics at High Energy,''
Camb. Monogr. Part. Phys. Nucl. Phys. Cosmol. \textbf{33}, 1-350 (2012)
Oxford University Press, 2013,
ISBN 978-1-009-29144-6, 978-1-009-29141-5, 978-1-009-29142-2, 978-0-521-11257-4, 978-1-139-55768-9
doi:10.1017/9781009291446
%330 citations counted in INSPIRE as of 20 May 2025

%\cite{Ivanov:2004ax}
\bibitem{Ivanov:2004ax}
I.~P.~Ivanov, N.~N.~Nikolaev and A.~A.~Savin,
%``Diffractive vector meson production at HERA: From soft to hard QCD,''
Phys. Part. Nucl. \textbf{37}, 1-85 (2006)
doi:10.1134/S1063779606010011
[arXiv:hep-ph/0501034 [hep-ph]].
%168 citations counted in INSPIRE as of 05 May 2025

%\cite{Rothkopf:2019ipj}
\bibitem{Rothkopf:2019ipj}
A.~Rothkopf,
%``Heavy Quarkonium in Extreme Conditions,''
Phys. Rept. \textbf{858}, 1-117 (2020)
doi:10.1016/j.physrep.2020.02.006
[arXiv:1912.02253 [hep-ph]].
%213 citations counted in INSPIRE as of 02 May 2025

%\cite{Lepage:1980fj}
\bibitem{Lepage:1980fj}
G.~P.~Lepage and S.~J.~Brodsky,
%``Exclusive Processes in Perturbative Quantum Chromodynamics,''
Phys. Rev. D \textbf{22}, 2157 (1980)
doi:10.1103/PhysRevD.22.2157
%4145 citations counted in INSPIRE as of 21 May 2025

%\cite{Gross:2022hyw}
\bibitem{Gross:2022hyw}
F.~Gross, E.~Klempt, S.~J.~Brodsky, A.~J.~Buras, V.~D.~Burkert, G.~Heinrich, K.~Jakobs, C.~A.~Meyer, K.~Orginos and M.~Strickland, \textit{et al.}
%``50 Years of Quantum Chromodynamics,''
Eur. Phys. J. C \textbf{83}, 1125 (2023)
doi:10.1140/epjc/s10052-023-11949-2
[arXiv:2212.11107 [hep-ph]].
%210 citations counted in INSPIRE as of 21 May 2025

%\cite{Vary:2009gt}
\bibitem{Vary:2009gt}
J.~P.~Vary, H.~Honkanen, J.~Li, P.~Maris, S.~J.~Brodsky, A.~Harindranath, G.~F.~de Teramond, P.~Sternberg, E.~G.~Ng and C.~Yang,
%``Hamiltonian light-front field theory in a basis function approach,''
Phys. Rev. C \textbf{81}, 035205 (2010)
doi:10.1103/PhysRevC.81.035205
[arXiv:0905.1411 [nucl-th]].
%213 citations counted in INSPIRE as of 20 May 2025

%\cite{Wiecki:2014ola}
\bibitem{Wiecki:2014ola}
P.~Wiecki, Y.~Li, X.~Zhao, P.~Maris and J.~P.~Vary,
%``Basis Light-Front Quantization Approach to Positronium,''
Phys. Rev. D \textbf{91}, no.10, 105009 (2015)
doi:10.1103/PhysRevD.91.105009
[arXiv:1404.6234 [nucl-th]].
%98 citations counted in INSPIRE as of 17 Apr 2025

%\cite{Li:2015zda}
\bibitem{Li:2015zda}
Y.~Li, P.~Maris, X.~Zhao and J.~P.~Vary,
%``Heavy Quarkonium in a Holographic Basis,''
Phys. Lett. B \textbf{758}, 118-124 (2016)
doi:10.1016/j.physletb.2016.04.065
[arXiv:1509.07212 [hep-ph]].
%143 citations counted in INSPIRE as of 02 May 2025

%\cite{Tang:2018myz}
\bibitem{Tang:2018myz}
S.~Tang, Y.~Li, P.~Maris and J.~P.~Vary,
%``$B_c$ mesons and their properties on the light front,''
Phys. Rev. D \textbf{98}, no.11, 114038 (2018)
doi:10.1103/PhysRevD.98.114038
[arXiv:1810.05971 [nucl-th]].
%58 citations counted in INSPIRE as of 14 May 2025

%\cite{Tang:2019gvn}
\bibitem{Tang:2019gvn}
S.~Tang, Y.~Li, P.~Maris and J.~P.~Vary,
%``Heavy-light mesons on the light front,''
Eur. Phys. J. C \textbf{80}, no.6, 522 (2020)
doi:10.1140/epjc/s10052-020-8081-9
[arXiv:1912.02088 [nucl-th]].
%49 citations counted in INSPIRE as of 02 May 2025

%\cite{Qian:2020utg}
\bibitem{Qian:2020utg}
W.~Qian, S.~Jia, Y.~Li and J.~P.~Vary,
%``Light mesons within the basis light-front quantization framework,''
Phys. Rev. C \textbf{102}, no.5, 055207 (2020)
doi:10.1103/PhysRevC.102.055207
[arXiv:2005.13806 [nucl-th]].
%40 citations counted in INSPIRE as of 20 May 2025

%\cite{Jia:2018ary}
\bibitem{Jia:2018ary}
S.~Jia and J.~P.~Vary,
%``Basis light front quantization for the charged light mesons with color singlet Nambu\textendash{}Jona-Lasinio interactions,''
Phys. Rev. C \textbf{99}, no.3, 035206 (2019)
doi:10.1103/PhysRevC.99.035206
[arXiv:1811.08512 [nucl-th]].
%83 citations counted in INSPIRE as of 13 May 2025

%\cite{Lan:2021wok}
\bibitem{Lan:2021wok}
J.~Lan \textit{et al.} [BLFQ],
%``Light mesons with one dynamical gluon on the light front,''
Phys. Lett. B \textbf{825}, 136890 (2022)
doi:10.1016/j.physletb.2022.136890
[arXiv:2106.04954 [hep-ph]].
%60 citations counted in INSPIRE as of 21 May 2025

%\cite{Mondal:2019jdg}
\bibitem{Mondal:2019jdg}
C.~Mondal, S.~Xu, J.~Lan, X.~Zhao, Y.~Li, D.~Chakrabarti and J.~P.~Vary,
%``Proton structure from a light-front Hamiltonian,''
Phys. Rev. D \textbf{102}, no.1, 016008 (2020)
doi:10.1103/PhysRevD.102.016008
[arXiv:1911.10913 [hep-ph]].
%81 citations counted in INSPIRE as of 20 May 2025

%\cite{Xu:2021wwj}
\bibitem{Xu:2021wwj}
S.~Xu \textit{et al.} [BLFQ],
%``Nucleon structure from basis light-front quantization,''
Phys. Rev. D \textbf{104}, no.9, 094036 (2021)
doi:10.1103/PhysRevD.104.094036
[arXiv:2108.03909 [hep-ph]].
%72 citations counted in INSPIRE as of 21 May 2025

%\cite{Xu:2024sjt}
\bibitem{Xu:2024sjt}
S.~Xu, Y.~Liu, C.~Mondal, J.~Lan, X.~Zhao, Y.~Li and J.~P.~Vary,
%``Towards a first principles light-front Hamiltonian for the nucleon,''
[arXiv:2408.11298 [hep-ph]].
%10 citations counted in INSPIRE as of 20 May 2025

%\cite{Kuang:2022vdy}
\bibitem{Kuang:2022vdy}
Z.~Kuang \textit{et al.} [BLFQ],
%``All-charm tetraquark in front form dynamics,''
Phys. Rev. D \textbf{105}, no.9, 094028 (2022)
doi:10.1103/PhysRevD.105.094028
[arXiv:2201.06428 [hep-ph]].
%31 citations counted in INSPIRE as of 14 May 2025

%\cite{Li:2017mlw}
\bibitem{Li:2017mlw}
Y.~Li, P.~Maris and J.~P.~Vary,
%``Quarkonium as a relativistic bound state on the light front,''
Phys. Rev. D \textbf{96}, 016022 (2017)
doi:10.1103/PhysRevD.96.016022
[arXiv:1704.06968 [hep-ph]].
%129 citations counted in INSPIRE as of 14 May 2025

%\cite{Nair:2024fit}
\bibitem{Nair:2024fit}
S.~Nair \textit{et al.} [BLFQ],
%``Gravitational form factors and mechanical properties of quarks in protons: A basis light-front quantization approach,''
Phys. Rev. D \textbf{110}, no.5, 056027 (2024)
doi:10.1103/PhysRevD.110.056027
[arXiv:2403.11702 [hep-ph]].
%13 citations counted in INSPIRE as of 20 May 2025

%\cite{Xu:2024hfx}
\bibitem{Xu:2024hfx}
S.~Xu, X.~Cao, T.~Hu, Y.~Li, X.~Zhao and J.~P.~Vary,
%``Gravitational form factors of charmonia,''
Phys. Rev. D \textbf{109}, no.11, 114024 (2024)
doi:10.1103/PhysRevD.109.114024
[arXiv:2404.06259 [hep-ph]].
%6 citations counted in INSPIRE as of 19 May 2025

%\cite{Hu:2024edc}
\bibitem{Hu:2024edc}
T.~Hu, X.~Cao, S.~Xu, Y.~Li, X.~Zhao and J.~P.~Vary,
%``Gravitational form factor D of charmonium from shear stress,''
Phys. Rev. D \textbf{111}, no.7, 074031 (2025)
doi:10.1103/PhysRevD.111.074031
[arXiv:2408.09689 [hep-ph]].
%3 citations counted in INSPIRE as of 19 May 2025

%\cite{Li:2018uif}
\bibitem{Li:2018uif}
M.~Li, Y.~Li, P.~Maris and J.~P.~Vary,
%``Radiative transitions between $0^{-+}$ and $1^{--}$ heavy quarkonia on the light front,''
Phys. Rev. D \textbf{98}, no.3, 034024 (2018)
doi:10.1103/PhysRevD.98.034024
[arXiv:1803.11519 [hep-ph]].
%41 citations counted in INSPIRE as of 17 Apr 2025

%\cite{Tang:2020org}
\bibitem{Tang:2020org}
S.~Tang, S.~Jia, P.~Maris and J.~P.~Vary,
%``Semileptonic decay of Bc to \ensuremath{\eta}c and J/\ensuremath{\psi} on the light front,''
Phys. Rev. D \textbf{104}, no.1, 016002 (2021)
doi:10.1103/PhysRevD.104.016002
[arXiv:2011.05454 [hep-ph]].
%9 citations counted in INSPIRE as of 17 Apr 2025

%\cite{Li:2021ejv}
\bibitem{Li:2021ejv}
Y.~Li, M.~Li and J.~P.~Vary,
%``Two-photon transitions of charmonia on the light front,''
Phys. Rev. D \textbf{105}, no.7, L071901 (2022)
doi:10.1103/PhysRevD.105.L071901
[arXiv:2111.14178 [hep-ph]].
%18 citations counted in INSPIRE as of 15 May 2025

%\cite{Wang:2023nhb}
\bibitem{Wang:2023nhb}
Z.~Wang, M.~Li, Y.~Li and J.~P.~Vary,
%``Shedding light on charmonium,''
Phys. Rev. D \textbf{109}, no.3, 3 (2024)
doi:10.1103/PhysRevD.109.L031902
[arXiv:2312.02604 [hep-ph]].
%4 citations counted in INSPIRE as of 20 May 2025

%\cite{Lan:2019vui}
\bibitem{Lan:2019vui}
J.~Lan, C.~Mondal, S.~Jia, X.~Zhao and J.~P.~Vary,
%``Parton Distribution Functions from a Light Front Hamiltonian and QCD Evolution for Light Mesons,''
Phys. Rev. Lett. \textbf{122}, no.17, 172001 (2019)
doi:10.1103/PhysRevLett.122.172001
[arXiv:1901.11430 [nucl-th]].
%109 citations counted in INSPIRE as of 20 May 2025

%\cite{Adhikari:2018umb}
\bibitem{Adhikari:2018umb}
L.~Adhikari, Y.~Li, M.~Li and J.~P.~Vary,
%``Form factors and generalized parton distributions of heavy quarkonia in basis light front quantization,''
Phys. Rev. C \textbf{99}, no.3, 035208 (2019)
doi:10.1103/PhysRevC.99.035208
[arXiv:1809.06475 [hep-ph]].
%32 citations counted in INSPIRE as of 15 May 2025

%\cite{Adhikari:2021jrh}
\bibitem{Adhikari:2021jrh}
L.~Adhikari \textit{et al.} [BLFQ],
%``Generalized parton distributions and spin structures of light mesons from a light-front Hamiltonian approach,''
Phys. Rev. D \textbf{104}, no.11, 114019 (2021)
doi:10.1103/PhysRevD.104.114019
[arXiv:2110.05048 [hep-ph]].
%27 citations counted in INSPIRE as of 20 May 2025

%\cite{Liu:2022fvl}
\bibitem{Liu:2022fvl}
Y.~Liu \textit{et al.} [BLFQ],
%``Angular momentum and generalized parton distributions for the proton with basis light-front quantization,''
Phys. Rev. D \textbf{105}, no.9, 094018 (2022)
doi:10.1103/PhysRevD.105.094018
[arXiv:2202.00985 [hep-ph]].
%32 citations counted in INSPIRE as of 20 May 2025

%\cite{Zhang:2023xfe}
\bibitem{Zhang:2023xfe}
Z.~Zhang \textit{et al.} [BLFQ],
%``Twist-3 generalized parton distribution for the proton from basis light-front quantization,''
Phys. Rev. D \textbf{109}, no.3, 034031 (2024)
doi:10.1103/PhysRevD.109.034031
[arXiv:2312.00667 [hep-th]].
%21 citations counted in INSPIRE as of 21 May 2025

%\cite{Lin:2023ezw}
\bibitem{Lin:2023ezw}
B.~Lin \textit{et al.} [BLFQ],
%``Generalized parton distributions of gluon in proton: A light-front quantization approach,''
Phys. Lett. B \textbf{847}, 138305 (2023)
doi:10.1016/j.physletb.2023.138305
[arXiv:2308.08275 [hep-ph]].
%21 citations counted in INSPIRE as of 21 May 2025

%\cite{Kaur:2023lun}
\bibitem{Kaur:2023lun}
S.~Kaur \textit{et al.} [BLFQ],
%``Spatial imaging of proton via leading-twist nonskewed GPDs with basis light-front quantization,''
Phys. Rev. D \textbf{109}, no.1, 014015 (2024)
doi:10.1103/PhysRevD.109.014015
[arXiv:2307.09869 [hep-ph]].
%23 citations counted in INSPIRE as of 21 May 2025

%\cite{Lin:2024ijo}
\bibitem{Lin:2024ijo}
B.~Lin \textit{et al.} [BLFQ],
%``Chiral-odd gluon generalized parton distributions in the proton: A light-front quantization approach,''
Phys. Lett. B \textbf{860}, 139153 (2025)
doi:10.1016/j.physletb.2024.139153
[arXiv:2408.09988 [hep-ph]].
%3 citations counted in INSPIRE as of 21 May 2025

%\cite{Liu:2024umn}
\bibitem{Liu:2024umn}
Y.~Liu \textit{et al.} [BLFQ],
%``Skewed generalized parton distributions of proton from basis light-front quantization,''
Phys. Lett. B \textbf{855}, 138809 (2024)
doi:10.1016/j.physletb.2024.138809
[arXiv:2403.05922 [hep-ph]].
%19 citations counted in INSPIRE as of 21 May 2025

%\cite{Xu:2022abw}
\bibitem{Xu:2022abw}
S.~Xu \textit{et al.} [BLFQ],
%``Quark and gluon spin and orbital angular momentum in the proton,''
Phys. Rev. D \textbf{108}, no.9, 094002 (2023)
doi:10.1103/PhysRevD.108.094002
[arXiv:2209.08584 [hep-ph]].
%42 citations counted in INSPIRE as of 21 May 2025

%\cite{Xu:2023nqv}
\bibitem{Xu:2023nqv}
S.~Xu \textit{et al.} [BLFQ],
%``Quark and gluon spin and orbital angular momentum in the proton,''
Phys. Rev. D \textbf{108}, no.9, 094002 (2023)
doi:10.1103/PhysRevD.108.094002
[arXiv:2209.08584 [hep-ph]].
%42 citations counted in INSPIRE as of 21 May 2025

%\cite{Hu:2022ctr}
\bibitem{Hu:2022ctr}
Z.~Hu \textit{et al.} [BLFQ],
%``Transverse momentum structure of proton within the basis light-front quantization framework,''
Phys. Lett. B \textbf{833}, 137360 (2022)
doi:10.1016/j.physletb.2022.137360
[arXiv:2205.04714 [hep-ph]].
%29 citations counted in INSPIRE as of 20 May 2025

%\cite{Zhu:2023lst}
\bibitem{Zhu:2023lst}
Z.~Zhu \textit{et al.} [BLFQ],
%``Transverse structure of the pion beyond leading twist with basis light-front quantization,''
Phys. Lett. B \textbf{839}, 137808 (2023)
doi:10.1016/j.physletb.2023.137808
[arXiv:2301.12994 [hep-ph]].
%24 citations counted in INSPIRE as of 20 May 2025

%\cite{Zhu:2024awq}
\bibitem{Zhu:2024awq}
Z.~Zhu \textit{et al.} [BLFQ],
%``Transverse structure of the proton beyond leading twist: A light-front Hamiltonian approach,''
Phys. Lett. B \textbf{855}, 138829 (2024)
doi:10.1016/j.physletb.2024.138829
[arXiv:2404.13720 [hep-ph]].
%5 citations counted in INSPIRE as of 21 May 2025

%\cite{Dyson:1949ha}
\bibitem{Dyson:1949ha}
F.~J.~Dyson,
%``The S matrix in quantum electrodynamics,''
Phys. Rev. \textbf{75}, 1736-1755 (1949)
doi:10.1103/PhysRev.75.1736
%1014 citations counted in INSPIRE as of 22 May 2025

%\cite{Schwinger:1951ex}
\bibitem{Schwinger:1951ex}
J.~S.~Schwinger,
%``On the Green's functions of quantized fields. 1.,''
Proc. Nat. Acad. Sci. \textbf{37}, 452-455 (1951)
doi:10.1073/pnas.37.7.452
%802 citations counted in INSPIRE as of 22 May 2025

%\cite{Schwinger:1951hq}
\bibitem{Schwinger:1951hq}
J.~S.~Schwinger,
%``On the Green's functions of quantized fields. 2.,''
Proc. Nat. Acad. Sci. \textbf{37}, 455-459 (1951)
doi:10.1073/pnas.37.7.455
%383 citations counted in INSPIRE as of 06 May 2025

%\cite{Maris:2003vk}
\bibitem{Maris:2003vk}
P.~Maris and C.~D.~Roberts,
%``Dyson-Schwinger equations: A Tool for hadron physics,''
Int. J. Mod. Phys. E \textbf{12}, 297-365 (2003)
doi:10.1142/S0218301303001326
[arXiv:nucl-th/0301049 [nucl-th]].
%652 citations counted in INSPIRE as of 07 May 2025

%\cite{Qin:2011dd}
\bibitem{Qin:2011dd}
S.~x.~Qin, L.~Chang, Y.~x.~Liu, C.~D.~Roberts and D.~J.~Wilson,
%``Interaction model for the gap equation,''
Phys. Rev. C \textbf{84}, 042202 (2011)
doi:10.1103/PhysRevC.84.042202
[arXiv:1108.0603 [nucl-th]].
%226 citations counted in INSPIRE as of 02 May 2025

%\cite{Roberts:2007ji}
\bibitem{Roberts:2007ji}
C.~D.~Roberts,
%``Hadron Properties and Dyson-Schwinger Equations,''
Prog. Part. Nucl. Phys. \textbf{61}, 50-65 (2008)
doi:10.1016/j.ppnp.2007.12.034
[arXiv:0712.0633 [nucl-th]].
%199 citations counted in INSPIRE as of 07 May 2025

%\cite{Cloet:2013jya}
\bibitem{Cloet:2013jya}
I.~C.~Cloet and C.~D.~Roberts,
%``Explanation and Prediction of Observables using Continuum Strong QCD,''
Prog. Part. Nucl. Phys. \textbf{77}, 1-69 (2014)
doi:10.1016/j.ppnp.2014.02.001
[arXiv:1310.2651 [nucl-th]].
%304 citations counted in INSPIRE as of 07 May 2025

%\cite{Eichmann:2016yit}
\bibitem{Eichmann:2016yit}
G.~Eichmann, H.~Sanchis-Alepuz, R.~Williams, R.~Alkofer and C.~S.~Fischer,
%``Baryons as relativistic three-quark bound states,''
Prog. Part. Nucl. Phys. \textbf{91}, 1-100 (2016)
doi:10.1016/j.ppnp.2016.07.001
[arXiv:1606.09602 [hep-ph]].
%427 citations counted in INSPIRE as of 19 May 2025

%\cite{Shi:2018zqd}
\bibitem{Shi:2018zqd}
C.~Shi and I.~C.~Clo\"et,
%``Intrinsic Transverse Motion of the Pion\textquoteright{}s Valence Quarks,''
Phys. Rev. Lett. \textbf{122}, no.8, 082301 (2019)
doi:10.1103/PhysRevLett.122.082301
[arXiv:1806.04799 [nucl-th]].
%33 citations counted in INSPIRE as of 17 Apr 2025

%\cite{Shi:2020pqe}
\bibitem{Shi:2020pqe}
C.~Shi, K.~Bednar, I.~C.~Clo\"et and A.~Freese,
%``Spatial and Momentum Imaging of the Pion and Kaon,''
Phys. Rev. D \textbf{101}, no.7, 074014 (2020)
doi:10.1103/PhysRevD.101.074014
[arXiv:2003.03037 [hep-ph]].
%29 citations counted in INSPIRE as of 21 May 2025

%\cite{Shi:2021taf}
\bibitem{Shi:2021taf}
C.~Shi, Y.~P.~Xie, M.~Li, X.~Chen and H.~S.~Zong,
%``Light front wave functions and diffractive electroproduction of vector mesons,''
Phys. Rev. D \textbf{104}, no.9, L091902 (2021)
doi:10.1103/PhysRevD.104.L091902
[arXiv:2101.09910 [hep-ph]].
%12 citations counted in INSPIRE as of 28 Apr 2025

%\cite{Shi:2021nvg}
\bibitem{Shi:2021nvg}
C.~Shi, M.~Li, X.~Chen and W.~Jia,
%``Ground state pseudoscalar mesons on the light front: From the light to heavy sector,''
Phys. Rev. D \textbf{104}, no.9, 094016 (2021)
doi:10.1103/PhysRevD.104.094016
[arXiv:2108.10625 [hep-ph]].
%9 citations counted in INSPIRE as of 28 Apr 2025

%\cite{Shi:2022erw}
\bibitem{Shi:2022erw}
C.~Shi, J.~Li, M.~Li, X.~Chen and W.~Jia,
%``Transverse momentum distributions of valence quarks in light and heavy vector mesons,''
Phys. Rev. D \textbf{106}, no.1, 014026 (2022)
doi:10.1103/PhysRevD.106.014026
[arXiv:2205.02757 [hep-ph]].
%15 citations counted in INSPIRE as of 15 May 2025

%\cite{Shi:2023oll}
\bibitem{Shi:2023oll}
C.~Shi, J.~Li, P.~L.~Yin and W.~Jia,
%``Unpolarized generalized parton distributions of light and heavy vector mesons,''
Phys. Rev. D \textbf{107}, no.7, 074009 (2023)
doi:10.1103/PhysRevD.107.074009
[arXiv:2302.02388 [hep-ph]].
%11 citations counted in INSPIRE as of 20 May 2025

%\cite{Kou:2023ady}
\bibitem{Kou:2023ady}
W.~Kou, C.~Shi, X.~Chen and W.~Jia,
%``Transverse momentum dependent parton distributions of pion at leading twist,''
Phys. Rev. D \textbf{108}, no.3, 036021 (2023)
doi:10.1103/PhysRevD.108.036021
[arXiv:2304.09814 [hep-ph]].
%10 citations counted in INSPIRE as of 17 Apr 2025

%\cite{Shi:2024laj}
\bibitem{Shi:2024laj}
C.~Shi, P.~Liu, Y.~L.~Du and W.~Jia,
%``Heavy flavor-asymmetric pseudoscalar mesons on the light front,''
Phys. Rev. D \textbf{110}, no.9, 094010 (2024)
doi:10.1103/PhysRevD.110.094010
[arXiv:2409.05098 [hep-ph]].
%6 citations counted in INSPIRE as of 15 May 2025

%\cite{Brodsky:1997de}
\bibitem{Brodsky:1997de}
S.~J.~Brodsky, H.~C.~Pauli and S.~S.~Pinsky,
%``Quantum chromodynamics and other field theories on the light cone,''
Phys. Rept. \textbf{301}, 299-486 (1998)
doi:10.1016/S0370-1573(97)00089-6
[arXiv:hep-ph/9705477 [hep-ph]].
%1636 citations counted in INSPIRE as of 20 May 2025



%\cite{Leitao:2017esb, Lan:2024ais}
\bibitem{Leitao:2017esb}
S.~Leit\~ao, Y.~Li, P.~Maris, M.~T.~Pe\~na, A.~Stadler, J.~P.~Vary and E.~P.~Biernat,
%``Comparison of two Minkowski-space approaches to heavy quarkonia,''
Eur. Phys. J. C \textbf{77}, no.10, 696 (2017)
doi:10.1140/epjc/s10052-017-5248-0
[arXiv:1705.06178 [hep-ph]].
%29 citations counted in INSPIRE as of 01 Jun 2025

%\cite{Lan:2024ais}
\bibitem{Lan:2024ais}
J.~Lan, C.~Mondal, X.~Zhao, T.~Frederico and J.~P.~Vary,
%``Gluonic contributions to the pion parton distribution functions,''
[arXiv:2406.18878 [hep-ph]].
%2 citations counted in INSPIRE as of 01 Jun 2025


%\cite{Ji:2003yj}
\bibitem{Ji:2003yj}
X.~d.~Ji, J.~P.~Ma and F.~Yuan,
%``Classification and asymptotic scaling of hadrons' light cone wave function amplitudes,''
Eur. Phys. J. C \textbf{33}, 75-90 (2004)
doi:10.1140/epjc/s2003-01563-y
[arXiv:hep-ph/0304107 [hep-ph]].
%68 citations counted in INSPIRE as of 30 Apr 2025

%\cite{Leitner:2010nx}
\bibitem{Leitner:2010nx}
O.~Leitner, J.~F.~Mathiot and N.~A.~Tsirova,
%``The Pion wave function in covariant light-front dynamics,''
Eur. Phys. J. A \textbf{47}, 17 (2011)
doi:10.1140/epja/i2011-11017-4
[arXiv:1009.5484 [hep-ph]].
%15 citations counted in INSPIRE as of 17 Apr 2025

%\cite{Jaus:1999zv}
\bibitem{Jaus:1999zv}
W.~Jaus,
%``Covariant analysis of the light front quark model,''
Phys. Rev. D \textbf{60}, 054026 (1999)
doi:10.1103/PhysRevD.60.054026
%175 citations counted in INSPIRE as of 06 May 2025

%\cite{Cheng:2003sm}
\bibitem{Cheng:2003sm}
H.~Y.~Cheng, C.~K.~Chua and C.~W.~Hwang,
%``Covariant light front approach for s wave and p wave mesons: Its application to decay constants and form-factors,''
Phys. Rev. D \textbf{69}, 074025 (2004)
doi:10.1103/PhysRevD.69.074025
[arXiv:hep-ph/0310359 [hep-ph]].
%402 citations counted in INSPIRE as of 12 May 2025

%\cite{Wang:2008xt}
\bibitem{Wang:2008xt}
W.~Wang, Y.~L.~Shen and C.~D.~Lu,
%``Covariant Light-Front Approach for B(c) transition form factors,''
Phys. Rev. D \textbf{79}, 054012 (2009)
doi:10.1103/PhysRevD.79.054012
[arXiv:0811.3748 [hep-ph]].
%162 citations counted in INSPIRE as of 16 May 2025

%\cite{Verma:2011yw}
\bibitem{Verma:2011yw}
R.~C.~Verma,
%``Decay constants and form factors of s-wave and p-wave mesons in the covariant light-front quark model,''
J. Phys. G \textbf{39}, 025005 (2012)
doi:10.1088/0954-3899/39/2/025005
[arXiv:1103.2973 [hep-ph]].
%161 citations counted in INSPIRE as of 22 May 2025

%\cite{Choi:2004ww}
\bibitem{Choi:2004ww}
H.~M.~Choi and C.~R.~Ji,
%``Electromagnetic structure of the rho meson in the light front quark model,''
Phys. Rev. D \textbf{70}, 053015 (2004)
doi:10.1103/PhysRevD.70.053015
[arXiv:hep-ph/0402114 [hep-ph]].
%90 citations counted in INSPIRE as of 20 May 2025

%\cite{Wang:2024cyi}
\bibitem{Wang:2024cyi}
S.~Y.~Wang, Y.~Y.~Yang, Z.~J.~Sun, H.~Yang, P.~Li and Z.~Q.~Zhang,
%``Semileptonic and nonleptonic decays of ${\boldsymbol B^{*}_{\boldsymbol {u,d,s}} }$ in the covariant light-front approach*,''
Chin. Phys. C \textbf{48}, no.12, 123102 (2024)
doi:10.1088/1674-1137/ad7247
[arXiv:2410.09672 [hep-ph]].
%2 citations counted in INSPIRE as of 17 Apr 2025

%\cite{Arifi:2024mff}
\bibitem{Arifi:2024mff}
A.~J.~Arifi, L.~Happ, S.~Ohno and M.~Oka,
%``Structure of heavy mesons in the light-front quark model,''
Phys. Rev. D \textbf{110}, no.1, 014020 (2024)
doi:10.1103/PhysRevD.110.014020
[arXiv:2401.07933 [hep-ph]].
%9 citations counted in INSPIRE as of 01 May 2025

%\cite{Li:2022mlg}
\bibitem{Li:2022mlg}
Y.~Li, P.~Maris and J.~P.~Vary,
%``Chiral sum rule on the light front and the 3D image of the pion,''
Phys. Lett. B \textbf{836}, 137598 (2023)
doi:10.1016/j.physletb.2022.137598
[arXiv:2203.14447 [hep-th]].
%7 citations counted in INSPIRE as of 17 Apr 2025



\bibitem{Li:charmoniumData}
Li, Yang (2019),
“Heavy quarkonium light front wave functions from basis light-front quantization with a running coupling”,
Mendeley Data, V2,
doi: 10.17632/cjs4ykv8cv.2

\bibitem{Shi:charmoniumData}
Shi, C. (2025). Light-front wave functions of the etaC meson from Dyson-Schwinger equations [Data set]. Zenodo. https://doi.org/10.5281/zenodo.18041666


%\cite{Miller:2010nz}
\bibitem{Miller:2010nz}
G.~A.~Miller,
%``Transverse Charge Densities,''
Ann. Rev. Nucl. Part. Sci. \textbf{60}, 1-25 (2010)
doi:10.1146/annurev.nucl.012809.104508
[arXiv:1002.0355 [nucl-th]].
%160 citations counted in INSPIRE as of 17 Apr 2025

%\cite{Miller:2018ybm}
\bibitem{Miller:2018ybm}
G.~A.~Miller,
%``Defining the proton radius: A unified treatment,''
Phys. Rev. C \textbf{99}, no.3, 035202 (2019)
doi:10.1103/PhysRevC.99.035202
[arXiv:1812.02714 [nucl-th]].
%111 citations counted in INSPIRE as of 12 May 2025

%\cite{Jaffe:2020ebz}
\bibitem{Jaffe:2020ebz}
R.~L.~Jaffe,
%``Ambiguities in the definition of local spatial densities in light hadrons,''
Phys. Rev. D \textbf{103}, no.1, 016017 (2021)
doi:10.1103/PhysRevD.103.016017
[arXiv:2010.15887 [hep-ph]].
%60 citations counted in INSPIRE as of 12 May 2025

%\cite{Drell:1969km}
\bibitem{Drell:1969km}
S.~D.~Drell and T.~M.~Yan,
%``Connection of Elastic Electromagnetic Nucleon Form-Factors at Large Q**2 and Deep Inelastic Structure Functions Near Threshold,''
Phys. Rev. Lett. \textbf{24}, 181-185 (1970)
doi:10.1103/PhysRevLett.24.181
%799 citations counted in INSPIRE as of 17 Apr 2025

%\cite{West:1970av}
\bibitem{West:1970av}
G.~B.~West,
%``Phenomenological model for the electromagnetic structure of the proton,''
Phys. Rev. Lett. \textbf{24}, 1206-1209 (1970)
doi:10.1103/PhysRevLett.24.1206
%519 citations counted in INSPIRE as of 17 Apr 2025

%\cite{Brodsky:1998hn}
\bibitem{Brodsky:1998hn}
S.~J.~Brodsky and D.~S.~Hwang,
%``Exact light cone wave function representation of matrix elements of electroweak currents,''
Nucl. Phys. B \textbf{543}, 239-252 (1999)
doi:10.1016/S0550-3213(98)00807-4
[arXiv:hep-ph/9806358 [hep-ph]].
%146 citations counted in INSPIRE as of 06 May 2025

%\cite{Dudek:2006ej}
\bibitem{Dudek:2006ej}
J.~J.~Dudek, R.~G.~Edwards and D.~G.~Richards,
%``Radiative transitions in charmonium from lattice QCD,''
Phys. Rev. D \textbf{73}, 074507 (2006)
doi:10.1103/PhysRevD.73.074507
[arXiv:hep-ph/0601137 [hep-ph]].
%229 citations counted in INSPIRE as of 15 May 2025

%\cite{Brodsky:2000ii}
\bibitem{Brodsky:2000ii}
S.~J.~Brodsky, D.~S.~Hwang, B.~Q.~Ma and I.~Schmidt,
%``Light cone representation of the spin and orbital angular momentum of relativistic composite systems,''
Nucl. Phys. B \textbf{593}, 311-335 (2001)
doi:10.1016/S0550-3213(00)00626-X
[arXiv:hep-th/0003082 [hep-th]].
%383 citations counted in INSPIRE as of 22 Apr 2025

%\cite{Cao:2023ohj}
\bibitem{Cao:2023ohj}
X.~Cao, Y.~Li and J.~P.~Vary,
%``Forces inside a strongly-coupled scalar nucleon,''
Phys. Rev. D \textbf{108}, no.5, 056026 (2023)
doi:10.1103/PhysRevD.108.056026
[arXiv:2308.06812 [hep-ph]].
%15 citations counted in INSPIRE as of 19 May 2025

%\cite{Cao:2024rul}
\bibitem{Cao:2024rul}
X.~Cao, S.~Xu, Y.~Li, G.~Chen, X.~Zhao, V.~A.~Karmanov and J.~P.~Vary,
%``Energy momentum tensor on and off the light cone: exposition with scalar Yukawa theory,''
JHEP \textbf{07}, 095 (2024)
doi:10.1007/JHEP07(2024)095
[arXiv:2405.06896 [hep-ph]].
%7 citations counted in INSPIRE as of 12 May 2025

%\cite{Cao:2024fto}
\bibitem{Cao:2024fto}
X.~Cao, Y.~Li and J.~P.~Vary,
%``Dissecting a strongly coupled scalar nucleon,''
Phys. Rev. D \textbf{110}, no.7, 076025 (2024)
doi:10.1103/PhysRevD.110.076025
[arXiv:2408.09535 [hep-ph]].
%3 citations counted in INSPIRE as of 28 Apr 2025

%\cite{Sultan:2024hep}
\bibitem{Sultan:2024hep}
M.~A.~Sultan, Z.~Xing, K.~Raya, A.~Bashir and L.~Chang,
%``Gravitational form factors of pseudoscalar mesons in a contact interaction,''
Phys. Rev. D \textbf{110}, no.5, 054034 (2024)
doi:10.1103/PhysRevD.110.054034
[arXiv:2407.10437 [hep-ph]].
%8 citations counted in INSPIRE as of 05 May 2025

%\cite{Davies:2010ip}
\bibitem{Davies:2010ip}
C.~T.~H.~Davies, C.~McNeile, E.~Follana, G.~P.~Lepage, H.~Na and J.~Shigemitsu,
%``Update: Precision $D_s$ decay constant from full lattice QCD using very fine lattices,''
Phys. Rev. D \textbf{82}, 114504 (2010)
doi:10.1103/PhysRevD.82.114504
[arXiv:1008.4018 [hep-lat]].
%263 citations counted in INSPIRE as of 08 May 2025

%\cite{Babiarz:2019sfa}
\bibitem{Babiarz:2019sfa}
I.~Babiarz, V.~P.~Goncalves, R.~Pasechnik, W.~Sch\"afer and A.~Szczurek,
%``${\gamma^* \gamma^* \to \eta_c (1S,2S)}$ transition form factors for spacelike photons,''
Phys. Rev. D \textbf{100}, no.5, 054018 (2019)
doi:10.1103/PhysRevD.100.054018
[arXiv:1908.07802 [hep-ph]].
%27 citations counted in INSPIRE as of 17 Apr 2025

%%\cite{Chen:2016bpj}
%\bibitem{Chen:2016bpj}
%J.~Chen, M.~Ding, L.~Chang and Y.~x.~Liu,
%%``Two Photon Transition Form Factor of $\bar{c}c $ Quarkonia,''
%Phys. Rev. D \textbf{95}, no.1, 016010 (2017)
%doi:10.1103/PhysRevD.95.016010
%[arXiv:1611.05960 [nucl-th]].
%%35 citations counted in INSPIRE as of 07 May 2025

\end{thebibliography}
\end{document}